\documentclass[acmsmall,screen]{acmart}

\AtBeginDocument{%
  }

\setcopyright{acmlicensed}
\copyrightyear{2024}
\acmYear{2024}
\acmDOI{XXXXXXX.XXXXXXX}

\acmYear{2024}
\acmDOI{XXXXXXX.XXXXXXX}

\acmJournal{TOMM}
\acmISBN{978-1-4503-XXXX-X/18/06}

\usepackage{latexsym}
\usepackage{hyperref}
\usepackage{booktabs}
\usepackage{multirow}
\usepackage{amsmath}
\usepackage{xparse}
\usepackage{caption}
\usepackage{subcaption}
\usepackage{graphicx}
\usepackage{algorithm}
\usepackage{algpseudocode}
\usepackage{colortbl}
\usepackage{tabularx}

\newcolumntype{P}[1]{>{\centering\arraybackslash}p{#1}}
\definecolor{ForestGreen}{rgb}{0.0, 0.5, 0.0} 

\begin{document}

\title{PenSLR: Persian end-to-end Sign Language Recognition Using Ensembling}

\author{Amirparsa Salmankhah}
\authornotemark[1]
\email{amirparsa.s@aut.ac.ir}
\orcid{0009-0002-4195-5164}
\affiliation{
  \institution{Amirkabir University of Technology (Tehran Polytechnic)}
  \streetaddress{P.O.Box: 15875-4413}
  \department{Department of Computer Engineering}
  \city{Tehran}
  \country{Iran}
  \postcode{15875-4413}
}

\author{Amirreza Rajabi}
\authornote{These authors contributed equally to this research.}
\orcid{0009-0001-6298-6667}
\email{dr.mrajabi.mr@aut.ac.ir}
\affiliation{
  \institution{Amirkabir University of Technology (Tehran Polytechnic)}
  \streetaddress{P.O.Box: 15875-4413}
  \department{Department of Computer Engineering}
  \city{Tehran}
  \country{Iran}
  \postcode{15875-4413}
}

\author{Negin Kheirmand}
\authornotemark[1]
\email{neginkheirmand@gmail.com}
\orcid{0009-0003-7538-3819}
\affiliation{
  \institution{Amirkabir University of Technology (Tehran Polytechnic)}
  \streetaddress{P.O.Box: 15875-4413}
  \department{Department of Computer Engineering}
  \city{Tehran}
  \country{Iran}
  \postcode{15875-4413}
}

\author{Ali Fadaeimanesh}
\authornotemark[1]
\orcid{0009-0005-3442-3292}
\email{alifadaeimanesh@aut.ac.ir}
\affiliation{
  \institution{Amirkabir University of Technology (Tehran Polytechnic)}
  \streetaddress{P.O.Box: 15875-4413}
  \department{Department of Computer Engineering}
  \city{Tehran}
  \country{Iran}
  \postcode{15875-4413}
}

\author{Amirreza Tarabkhah}
\authornotemark[1]
\orcid{0009-0002-6732-8441}
\email{tarabkhah2@aut.ac.ir}
\affiliation{%
  \institution{Amirkabir University of Technology (Tehran Polytechnic)}
  \streetaddress{P.O.Box: 15875-4413}
  \department{Department of Computer Engineering}
  \city{Tehran}
  \country{Iran}
  \postcode{15875-4413}
}

\author{Amirreza Kazemzadeh}
\authornotemark[1]
\orcid{0009-0007-9390-2247}
\email{ar.kazemzade@gmail.com}
\affiliation{%
  \institution{Amirkabir University of Technology (Tehran Polytechnic)}
  \streetaddress{P.O.Box: 15875-4413}
  \department{Department of Computer Engineering}
  \city{Tehran}
  \country{Iran}
  \postcode{15875-4413}
}

\author{Hamed Farbeh}
\authornote{Hamed Farbeh is the corresponding author.}
\orcid{0000-0002-4204-9131}
\email{farbeh@aut.ac.ir}
\affiliation{
  \institution{Amirkabir University of Technology (Tehran Polytechnic)}
  \streetaddress{P.O.Box: 15875-4413}
  \department{Department of Computer Engineering}
  \city{Tehran}
  \country{Iran}
  \postcode{15875-4413}
}
\renewcommand{\shortauthors}{Salmankhah et al.}

\begin{abstract}
Sign Language Recognition (SLR) is a fast-growing field that aims to fill the communication gaps between the hearing-impaired and people without hearing loss. Existing solutions for Persian Sign Language (PSL) are limited to word-level interpretations, underscoring the need for more advanced and comprehensive solutions. Moreover, previous work on other languages mainly focuses on manipulating the neural network architectures or hardware configurations instead of benefiting from the aggregated results of multiple models. In this paper, we introduce PenSLR, a glove-based sign language system consisting of an Inertial Measurement Unit (IMU) and five flexible sensors powered by a deep learning framework capable of predicting variable-length sequences. We achieve this in an end-to-end manner by leveraging the Connectionist Temporal Classification (CTC) loss function, eliminating the need for segmentation of input signals. To further enhance its capabilities, we propose a novel ensembling technique by leveraging a multiple sequence alignment algorithm known as Star Alignment. Furthermore, we introduce a new PSL dataset, including 16 PSL signs with more than 3000 time-series samples in total. We utilize this dataset to evaluate the performance of our system based on four word-level and sentence-level metrics.
Our evaluations show that PenSLR achieves a remarkable word accuracy of 94.58\% and 96.70\% in subject-independent and subject-dependent setups, respectively.  These achievements are attributable to our ensembling algorithm, which not only boosts the word-level performance by 0.51\% and 1.32\% in the respective scenarios but also yields significant enhancements of 1.46\% and 4.00\%, respectively, in sentence-level accuracy. 
\end{abstract}





\maketitle

\section{Introduction}

There are more than 72 million users worldwide who make use of sign language on a daily basis. In Iran, the number of deaf and hard-of-hearing individuals is estimated to be over 3 million as of 2019 \cite{zei}. These people face communication problems, leading to social isolation, which can impact their quality of life. The two main solutions for communication between sign language users and the rest of the population are either through handwriting or an interpreter, which can sometimes be neither feasible in everyday scenarios nor fast or interactive. Moreover, sign languages can be diverse, with multiple variations existing in a single region or country. This could lead to difficulties in communication, even between sign language users.

Sign Language Recognition (SLR) has emerged as a rapidly developing field within the research community. It focuses on addressing the problem of recognizing sentence-level sign language glosses. SLR can help bridge communication gaps between deaf or hard-of-hearing individuals who use sign language and those who do not, enabling more inclusive interactions in various settings.
In understanding sign languages, Persian Sign Language (PSL), like others, consists of a combination of intricate finger positions, hand movements, and face mimics that collectively define its unique characteristics. These gestures fall into two categories: manual and non-manual markers. The former indicates the finger positions and trajectory of the hand throughout the gesture, and the latter refers to facial expressions or head motion. Given the critical role of manual and non-manual markers in SLR, any effective SLR system must incorporate as many of these features into its recognition process as possible. 

A number of approaches exist to tackle the SLR problem, each with its own set of pros and cons. The main methods are vision-based and wearable-based SLR systems. The vision-based approach consists of analyzing visual signals through the use of cameras or other visual sensors, and it has the advantage of considering both manual and non-manual markers \citep{17-karami2011persian,13-zhang2023c2st,19-zuo2022improving,20-hu2021global,14-mittal2019modified}. However, this approach also comes with its set of drawbacks, such as its undermining of privacy, complexity of the data gathering process, and sensitivity to lighting conditions. The second approach involves wearable-based systems that depend on a glove or any other type of wearable device attached to different parts of the hands or head. The embedded sensors provide the data that will be processed and translated into sign language glosses. A number of these methods make use of traditional machine learning approaches \citep{15-wu2016wearable,6-khomami2021persian,10-zhou2020sign,7-lu2023data,8-li2020movement} while others leverage the power of deep learning \citep{23-basiri2023dynamic,5-wen2021ai,24-liu2023wearable,3-delpreto2022wearable,11-zhang2019myosign,25-sharma2021continuous,26-zhang2022wearsign,27-wang2020hear} to recognize and translate sign language. Although these systems mostly come with the disadvantage of not taking into account the non-manual markers and depend on individual anatomy, their privacy-preserving nature, portability, and affordability make them suitable choices for SLR. Recent research in this field has also introduced a new approach based on wireless sensing \citep{16-jin2023smartasl, 22-jin2021sonicasl, 21-santhalingam2020mmasl, 18-shang2017robust}. These types of SLR systems depend on the analysis of acoustic or non-acoustic waves and usually have the advantage of less computational complexity and portability. However, these solutions may experience interference from external waves and underperform in environments where obstacles hinder wave propagation.

Researchers may face several challenges while developing an SLR system. Firstly, the input data is dependent on both spatial and temporal features. The former is due to the spatial nature of this task, while the latter originates from the complexity of glosses as well as the variations in sign language users' speed. Secondly, It can be challenging to distinguish between some sign language gestures due to their similarities in hand movements, finger positions, or facial expressions. Moreover, due to variations in the way that sign language users perform gestures, the trained model must generalize training data to maintain accuracy when input from a new user is fed into the model. This also must be taken into account when predicting a previously unseen sequence from a user in the training set. Lastly, the model must be capable of detecting the transitions between the glosses to achieve high accuracy when facing new sentences with much higher lengths than the training samples.

In this paper, we propose a glove-based SLR system to detect variable-length sentences from PSL using an end-to-end deep learning framework. We leverage a customized sign language glove with a low-cost Inertial Measurement Unit (IMU) attached to the back of the hand and five flexible sensors mounted on fingers. We collect two short-length and long-length datasets to prove the ability of our framework to generalize in longer unseen sentences. Our deep learning framework consists of two main parts. Firstly, we design a Convolutional Recurrent Neural Network (CRNN) to extract the spatio-temporal features from the given input sequence and predict the corresponding label. We achieve this by utilizing the widely-used Connectionist Temporal Classification (CTC) loss function \cite{graves2006ctc}, which allows us to optimize the alignments between the input sequences and the ground truth without any prior knowledge about the alignments or any segmentation scheme. Secondly, we propose a novel approach for ensembling that makes use of multiple trained models and performs a voting process to obtain the final result. Our ensembling algorithm utilizes Star Alignment, a popular multiple sequence alignment algorithm, to align the predicted sequences of the models, ensuring they are of equal length. It will then take a majority vote between the aligned sequences at each position to generate the final prediction. Our framework leads to benefits regarding unseen sentences, specifically longer ones. Furthermore, this scheme improves the system's robustness in adapting to new individuals. In addition, since the ensembling method does not rely on any particular characteristic of PSL, future work in this domain could take advantage of it to enhance their result. Ultimately, its applicability extends to other domains like gesture recognition, given its independence from linguistic attributes.

In order to assess the effectiveness of our system, we gathered a dataset comprising over 3000 samples from 16 commonly used PSL glosses. The dataset was collected with the help of five volunteers and includes sentences of up to three words from the selected glosses. Another dataset, containing 4 to 8-word sentences, was gathered with the aim of evaluating the model’s ability to recognize longer sentences. Since it is not a realistic scenario to train a new model for each new user, we conduct a subject-independent analysis to guarantee the practicality of our system. We compare the two best models obtained by an ablation study in terms of four word-level and sentence-level evaluation metrics. The results depict that our system achieves almost 94\% word-level accuracy in both datasets, showcasing its proficiency in detecting both short and long sequences. The model is capable of predicting the length of the sequences in approximately 95\% of samples, and a sentence-level exact match ratio of almost 88\% and 80\% is achieved for the short-length and long-length datasets, respectively. 

To sum up, our main contributions are summarized as follows:
\begin{itemize}
    \item We design a low-cost sign language glove using an IMU and five flexible sensors capable of capturing a wide range of finger bendings and hand movements.
    \item We collect two datasets containing short-length and long-length time-series data from PSL sentences containing 16 widely-used PSL glosses. We make our dataset publicly available\footnote{https://github.com/Persian-Sign-Language/PenSLR-dataset} in order to make it accessible to researchers in the field of SLR, especially the ones working on PSL.
    \item We develop a CRNN architecture with the ability to process variable-length signals and predict complete sign language sentences in an end-to-end fashion.
    \item We propose a brand new ensembling scheme using Star Alignment as its backbone, which is adaptable to other SLR or sequence-to-sequence tasks. 
\end{itemize}

The rest of the paper is organized as follows. Section \ref{sec:related-work} presents the previous work in the field of SLR and provides details about multiple sequence alignment algorithms related to our ensembling method. Section \ref{sec:dataset} describes our dataset as well as the data gathering process. Section \ref{sec:our-method} deeply investigates different parts of PenSLR, including the designed glove, the architecture of the model, and our ensembling scheme. In Section \ref{sec:eval}, the performance of the system is discussed using multiple experiments and evaluation metrics. Then, in Section \ref{sec:limit} we investigate the limitations of our system and explore possible avenues for future research. Finally, the conclusion of the paper is drawn in Section \ref{sec:conc}.

\section{Related Work} \label{sec:related-work}

In this section, we explore the related works in the field of sign language recognition and provide a brief explanation of multiple sequence alignment algorithms.

\subsection{Sign Language Recognition}

Two leading solutions exist to tackle the problem of sign language recognition (SLR) based on sensing technologies: vision-based and mobile/wearable solutions. Also, a recently emerging approach is based on wireless sensing.

\subsubsection{Vision-based Methods}

Vision-based methods usually utilize camera setup and visual signals, thus have the advantage of considering non-manual markers. An early work \cite{17-karami2011persian} in this area for PSL consists of the use of Discrete Wavelet Transform (DWT) for feature extraction and a Multi-layer Perceptron (MLP) neural network for categorizing sign language gestures. In another study, C2ST \cite{13-zhang2023c2st} takes into account the linguistic features of gloss sequences using a linguistic model. Later research in the vision field ensures spatial attention consistency using a keypoint-guided spatial attention module \cite{19-zuo2022improving}. Another work in this field also takes into account non-manual markers using a custom-developed Global-local enhancement network (GLE-Net) architecture \cite{20-hu2021global}. Moreover, \cite{14-mittal2019modified} suggests using a Leap Motion sensor to extract the coordinates of different parts of the hands. It also propose a modified version of Long Short Term Memory (LSTM), introducing a new reset gate that resets the memory of LSTM whenever a non-active situation is detected. 

\subsubsection{Wireless sensing-based Methods}

The following methods introduce wireless sensing-based SLR systems that use electromagnetic or acoustic waves to detect body movements. In \cite{18-shang2017robust}, a Wi-Fi receiver and two Wi-fi transmitters are used to collect data, and then a Kernel-based Support Vector Machine (SVM) model is used to classify gestures on different systems. Another work, mmASL \cite{21-santhalingam2020mmasl}, takes advantage of millimeter waves and a multi-task deep learning model to recognize American Sign Language (ASL) gestures. Additionally, SonicASL \cite{22-jin2021sonicasl}, a real-time gesture recognition system, identifies sign language through the use of earphones with built-in microphones and speakers, achieving high accuracy rates in both word and sentence recognition. In a later work, SmartASL \cite{16-jin2023smartasl} not only utilizes earbud signals but also employs IMU sensors to detect both manual and non-manual markers. The last two mentioned works employ CRNN architecture along with CTC loss function to perform sign language recognition in real-time.  

\subsubsection{Wearable-based Methods} \label{sec:rel-wearable}

Mobile/Wearable systems designed to recognize hand or body gestures are another approach to the SLR task. Although these solutions do not consider non-manual markers, they can help solve the range of failure points that the previously mentioned vision-based and wireless sensing-based solutions have. 

An early work in this field used customized bending sensors, accelerometers, and Hall effect sensors in order to detect gestures of digits (0 to 9) through logistic regression \cite{9-chouhan2014smart}. Furthermore, authors in \cite{1-muralidharan2022modelling} suggest a 5-bit representation of the words or sentences and assign a bit to the output of each flexible sensor on fingers to predict the gestures based on the perceived values. Another study involves using a combination of IMU and sEMG sensors to gather data \cite{15-wu2016wearable}. The sEMG sensors make it possible for them to automatically split the input signals into segments that are fed to an SVM classifier for the final prediction. Moreover, a study on PSL suggests using a similar sensor setting along with a KNN classifier and K-fold cross-validation to detect 20 commonly used isolated sign gestures \cite{6-khomami2021persian}. However, it did not provide a solution for recognizing PSL sentences. In \cite{10-zhou2020sign}, a yarn-based stretchable sensor was introduced, which is not only capable of capturing hand gestures but can also be attached to eyebrows and mouth. Then, an SVM model was used to perform the classification task. Authors in \cite{7-lu2023data} took a unique approach by storing predefined sentences in a database while harnessing a customized glove with an IMU and flexible sensors. During the testing phase, a novel DTW distance was proposed to find the nearest data in the database and predict the label of the new input signal. In \cite{8-li2020movement}, the authors suggested an approach to calculate the movement trajectories of gestures and took advantage of another customized DTW distance to classify the signals.

Later works leverage the power of deep learning to achieve better results and accomplish more complex tasks. A work on PSL \cite{23-basiri2023dynamic} collected a time-series dataset of 15 words with 600 samples in total. The collected data were virtually augmented to 30000 images using the State-Image approach, and a CNN network was proposed to train on these images. Similar to \cite{6-khomami2021persian}, this study lacked the ability to predict continuous sentences. Another study suggests using a sliding-window approach to segment the input signals and construct the predicted sentence by feeding the segments to a CNN-based neural network \cite{5-wen2021ai}. This enabled the model to achieve high results in detecting 50 words and 20 predefined sentences, as well as new sentences that could be constructed by different combinations of those words. Authors in \cite{24-liu2023wearable} utilized a similar CNN-based network with a customized glove containing an IMU and strain sensors to detect 48 Chinese Sign Language (CSL) gestures. They took advantage of multiple sliding windows with different lengths and aggregated their results for sentence-level prediction. In another study, a glove was designed using conductive knit fabric and an accelerometer \cite{3-delpreto2022wearable}. To generate the output, an LSTM network was proposed to train on a dataset containing 12 distinct classes from ASL using a sliding-window approach.

The most recent approaches involve using CRNN models along with CTC loss function or an extension of them involving attention-based networks. Myosign \cite{11-zhang2019myosign} suggests an armband capable of collecting data from different modalities, i.e., accelerometer, gyroscope, orientation, and EMG. They propose a CRNN network consisting of multimodal CNN and Bidirectional LSTM (BiLSTM) layers to build an end-to-end system for predicting ASL sentences. They achieve this through the use of CTC loss, which enables them to perform SLR without any prior knowledge of the alignments between input and output. In \cite{25-sharma2021continuous}, a transfer learning scheme was employed on a similar architecture as Myosign, enabling the system to converge faster while maintaining accuracy. Their wearable system consisted of multiple IMU sensors on both forearms. Building on Myosign, Wearsign \cite{26-zhang2022wearsign} introduces an encoder-decoder framework that takes advantage of the attention mechanism to translate the sign language glosses to the spoken text. Moreover, they utilized the back-translation technique to augment and extend their ASL dataset. Similarly, Hear Sign \cite{27-wang2020hear} proposes an attention-based encoder-decoder approach with a multi-channel CNN to train on the data collected by two MYO armbands effectively.

\subsection{Multiple Sequence Alignment}
Due to the increase of data in today's era, especially sequence data, various algorithms have been designed to align multiple sequences together. The application of these algorithms is mainly in the field of bioinformatics. For example, understanding the functional significance of genetic variations across diverse species counts as one of their primary applications. The task of aligning two sequences, pairwise sequence alignment, serves as the foundation for aligning multiple sequences. Pairwise alignment is primarily categorized into local alignment and global alignment. Local alignment focuses on identifying and aligning similar local regions, while global alignment involves aligning sequences in an end-to-end manner \cite{msa-developments}. In this section, our primary focus relies on global alignment because it is the approach we use in our proposed algorithm. 

The Needleman-Wunsch (NW) algorithm \cite{needleman} is a widely employed global alignment algorithm that leverages a dynamic programming approach to calculate pairwise sequence alignment, taking advantage of the optimal substructure of this problem. However, using this algorithm for multiple sequence alignment is not feasible due to its time complexity. In order to tackle this problem, two main strategies are suggested: star alignment and progressive alignment. Star alignment works by automatically selecting one of the sequences as the center and aggregating the results of pairwise alignment between the center and the other sequences to obtain the final alignment. This approach considerably reduces the time needed to compute the final alignments, thus making it a suitable choice for real-time tasks. On the other hand, progressive alignment algorithms, such as ClustalW \cite{clustalw}, are based on a guide tree that determines the order by which the sequences should get aligned. There are two main drawbacks to the progressive alignment algorithms. First, the final result is highly dependent on the quality of the guide tree. Second, they need more time to execute as hierarchical clustering is needed to construct the guide tree, and internal nodes of the tree require more complex operations for computing the alignments \cite{msa-developments}.

\section{Dataset} \label{sec:dataset}
We collected our own dataset for two reasons. First, as mentioned in Section \ref{sec:rel-wearable}, there was only one public PSL dataset \cite{23-basiri2023dynamic} available, which was neither large enough nor contained sentence-level data. Second, none of the previous work on PSL used our hardware settings, forcing us to collect new data to ensure the compatibility of our glove-based design with our dataset. Unlike many conducted studies in the area of SLR, which used separated word-level and sentence-level datasets to train the model, we collected a unified dataset containing word combinations (sentences) of up to three words. Sentences are random permutations of the selected PSL signs and do not necessarily convey meaning. Moreover, the reason for limiting the number of words in sentences was to show the capability of our Seq2Seq model to predict sentences containing more than three words without being trained on them.

\begin{figure*}[b]
    \centering
    \includegraphics[scale=0.4]{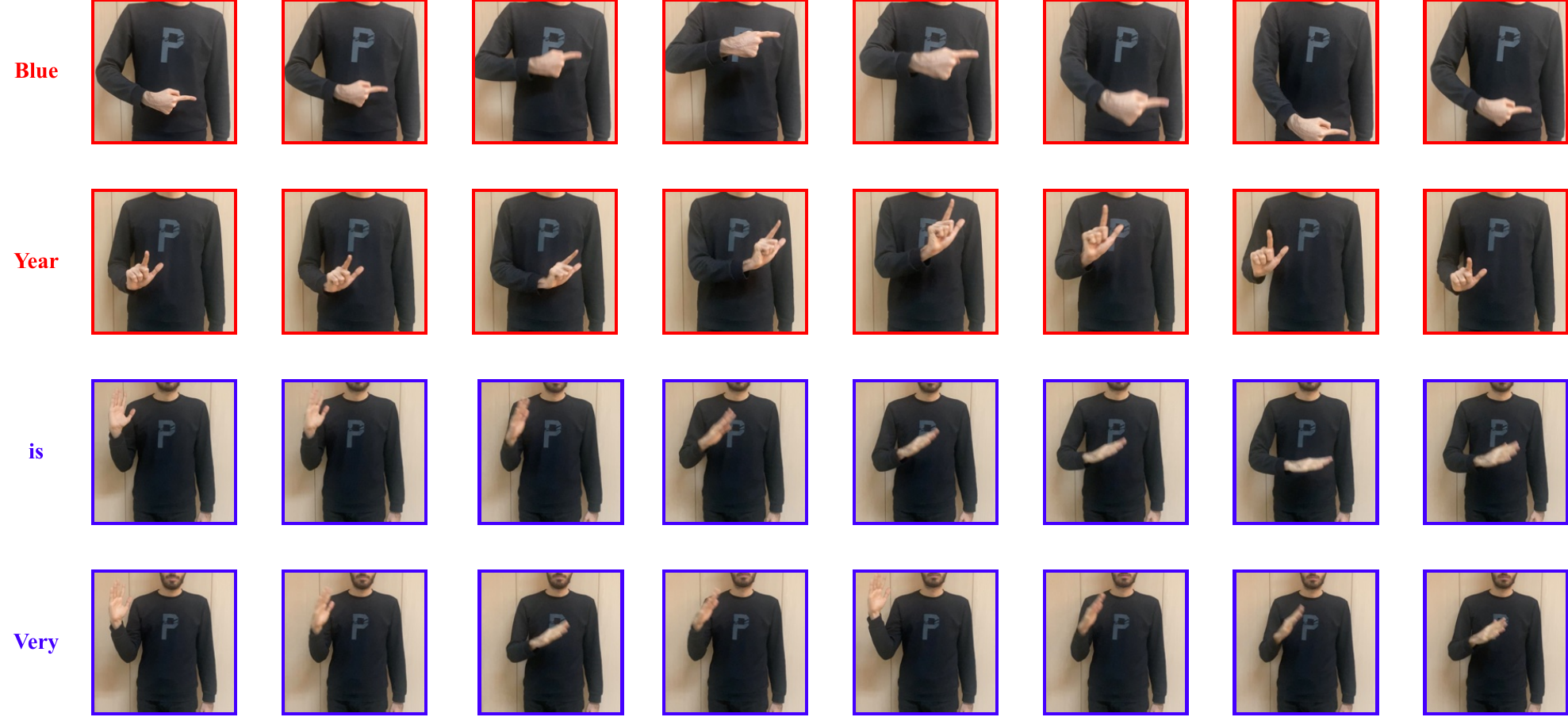}
    \caption{Illustration of step-by-step execution of two pairs of PSL glosses ("Blue", "Year") and ("is", "Very") belonging to two distinct similarity groups. The blue glosses have similar finger positions but different hand movements, while the red ones are examples of rotational gestures.}
    \label{fig:psl-moves}
\end{figure*}

To demonstrate the effectiveness of our model, we cherry-picked 16 words from PSL, categorizing them into five similarity groups. Table \ref{tab:psl-similarity-groups} demonstrates the groups and the relationship between the similar words. The words within the same tuple in groups 1 and 2 mutually share the characteristics of the group. That is, they share similarities in at least one aspect, either through hand movements or finger positions, making them indistinguishable without the concurrent use of both sensors. On the other hand, groups 3 to 5 feature the words that have fixed, dynamic, or rotational hand movements, respectively. As a result, achieving high accuracy in this setup could showcase the model’s proficiency in detecting and distinguishing minor variations between similar gestures. For clearer understanding, Figure \ref{fig:psl-moves} depicts the execution of two pairs of our PSL words distinguished by different colors. The red group shows a pair of  gestures ("Blue" and "Year") with rotational hand movements on different axes. The blue group illustrates gestures ("is" and "Very"), which share finger positions but have minor differences in hand movement.


\begin{table}[t]
    \resizebox{0.9\textwidth}{!}{%
     \begin{tabular}{P{0.5cm}|P{7.25cm}|P{11.25cm}}
     \toprule
          \textbf{ID} & \textbf{Group Characteristic} & \textbf{Members} \\
     \midrule
     1 & Same finger position but different hand movement & (Agreement, Disagreement) - (Yesterday, Father, Luck, Year) - (is, Very, Hopeful) \\
     2 & Same hand movement but different finger position & (Luck, Summer) \\
     3 & Fixed hand movement and finger position & Good - Agreement - Disgreement \\
     4 & Dynamic hand movement and finger position & Day - Forget - Mother \\
     5 & Rotational movements & Blue - Green - Year \\
     \bottomrule
     \end{tabular}}
\caption{Our selected PSL words categorized into five groups}
\label{tab:psl-similarity-groups}
\end{table}

We asked five volunteers to perform sign language gestures. Although recording hand-picked sentences multiple times by different volunteers is a common way to build a sign language dataset, we randomly generated the sentences in our dataset, ensuring no human bias is involved in the selection process. For our primary dataset (Dataset1-3), each volunteer recorded, on average, 100 one-word sentences, 300 two-word sentences, and 200 three-word sentences. Moreover, we asked each volunteer to record 20 extra sentences containing 4 to 8 words to evaluate the ability of our model in longer unseen sentences (Dataset4-8). Additionally, a GUI application was designed and implemented to increase the speed and ease of the recording process.

Each data in the dataset is a sequence of values received from the sensors at a rate of 100 Hz. IMU features include total acceleration, linear acceleration, gyroscope, and gravity acceleration in the X, Y, and Z axes. Combining these features with five features returned from flexible sensors forms 17 distinct features per data point. Therefore, every sample in the dataset is a time series data that can have different lengths depending on the execution time of each gesture. 

Table \ref{tab:average-length-of-gestures} depicts the average number of data points derived from sequences containing one to three words across different subjects. According to the table, firstly, the average amount of time to execute sentences with the same length is different for different subjects. Secondly, subjects show different behaviors for different sentence lengths and do not have a constant speed while performing gestures. For example, on average, subject 4 has performed single-word sentences in the least amount of time but exhibits the slowest pace for 3-word sentences. These variations pose a challenge for the model in detecting word transitions within sentences.

\begin{table}[t]
    \resizebox{0.75\textwidth}{!}{%
     \begin{tabular}{P{1.5cm}P{1.5cm}P{1.5cm}P{1.5cm}P{1.5cm}P{1.5cm}P{1.5cm}}
     \toprule
          & \textbf{Subject 1} &                                          \textbf{Subject 2} &                                          \textbf{Subject 3} &                                          \textbf{Subject 4} &
          \textbf{Subject 5} &
          \textbf{Average} \\
     \midrule
     \centering
     \textbf{Length 1}  & \textcolor{ForestGreen}{13.0} & 11.7 & 11.6 & \textcolor{red}{9.7} & 10.7 & 11.34 \\
     \textbf{Length 2}  & 19.0 & 19.1 & \textcolor{red}{18.2} & 19.8 & \textcolor{ForestGreen}{20.2} &  19.26\\
     \textbf{Length 3}  & \textcolor{red}{25.4} & 28.0 & 27.3 & \textcolor{ForestGreen}{30.2} & 26.5 & 27.48 \\
    
     \bottomrule
     \end{tabular}}
\caption{The average number of sample lengths in Dataset1-3 grouped by different subjects. The slowest and the fastest subjects in each row are denoted by green and red, respectively.}
\label{tab:average-length-of-gestures}
\end{table}


The dataset is intended to facilitate the research toward PSL or any other sign language variation. To ensure the reproducibility and accessibility of our work, we have made the dataset publicly available\footnote{https://github.com/Persian-Sign-Language/PenSLR-dataset}. We encourage other researchers, especially the ones working on PSL, to use our dataset to become familiar with the challenges that may arise while working with sign language datasets or to propose new architectures that could achieve better results than our model.

\algnewcommand{\LeftComment}[1]{\Statex \(\triangleright\) \textcolor{violet}{#1}}

\algnewcommand{\RightComment}[1]{\Comment{\textcolor{violet}{#1}}}

\section{Proposed Method} \label{sec:our-method}
In this section, we describe different parts of PenSLR. This includes the design of the sign language glove, the data collection process, preprocessing techniques, the deep learning framework, and the proposed ensembling method.

\subsection{Glove Design}

\begin{figure*}
    \centering
    \includegraphics[scale=0.2]{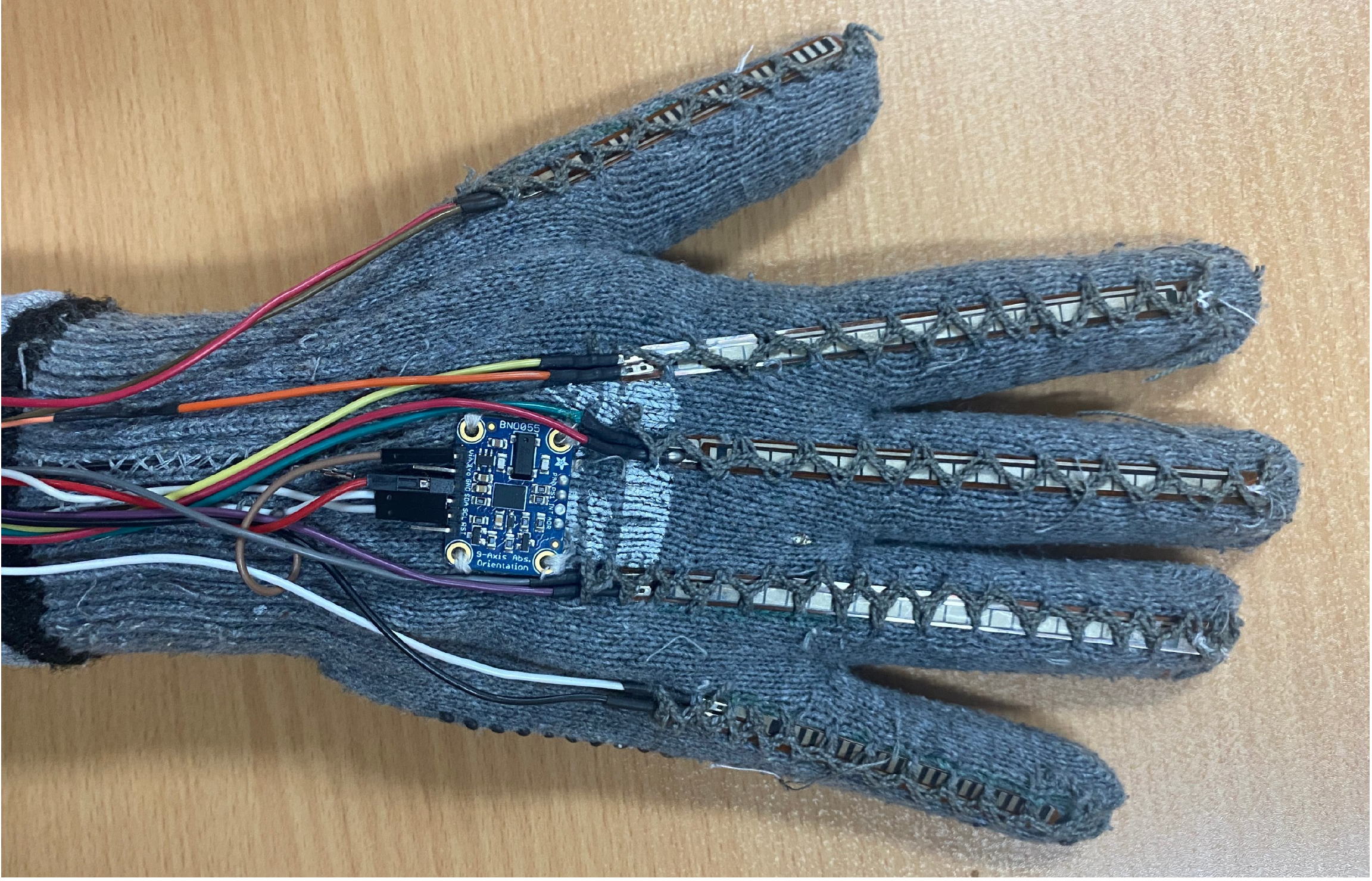}
    \caption{Our sign language glove equipped with an Adafruit BNO055 IMU mounted on the back of the hand and five flexible sensors on each finger.}
    \label{fig:glove-structure}
\end{figure*}

As depicted in Figure \ref{fig:glove-structure}, our designed glove incorporates two types of sensors: an IMU sensor mounted on the back of the hand and five flexible sensors attached to each finger. We utilized Arduino Mega to collect the sensors' data. The IMU sensor sends its data to the microcontroller via I2C protocol, whereas the flexible sensors achieve this by analog input pins.

One of the main goals of our research was to design a low-cost but accurate glove. As a result, we used the affordable Adafruit BNO055 IMU, which can deliver a variety of metrics, including acceleration, orientation, and gravity, along three distinct axes. Capturing these features enables us to distinguish between different hand movements based on orientation patterns. flexible sensors, on the other hand, are sensitive to bending; thus, their values change with the closing and opening of fingers. Considering the setup above, our glove is capable of covering a wide range of hand gestures while being cost-effective and easily repairable.

\subsection{Seq2Seq Model}
This section describes the specifics of our model, such as preprocessing steps, the model architecture and the loss function we use in our model.

\subsubsection{Preprocessing}
Since errors (caused by human or hardware factors) may affect the data collection process, it is necessary to identify outliers and remove them from the dataset before using them. To achieve this, we used the Interquartile Range (IQR) method. In this method, the first quartile $Q_{1,f}$ and the third quartile $Q_{3,f}$  of data are calculated for each feature and the difference between them $IQR_f$ is used to determine a range for non-outlier data. To be precise, when a data point falls outside the range of $[Q_{1,f} - 1.5 \cdot IQR_f, Q_{3,f} + 1.5 \cdot IQR_f]$ for at least one feature, the whole sample is considered as an outlier and will be removed from the dataset.

Data normalization was another step in our preprocessing pipeline. To normalize the data, we calculated the minimum value $x_{min, f}$ and the maximum value $x_{max, f}$ of each feature based on the training data. Then, we normalized the value of each feature $x_f$ in training and validation data as follows:
\begin{equation}
    x_f = \frac{x_f - x_{min, f}}{x_{max,f} - x_{min, f}}
\end{equation}
During the testing phase, the test data were normalized with respect to the values we calculated beforehand using the training dataset.
\begin{figure*}[!t]
    \centering
    \includegraphics[width=1\textwidth]{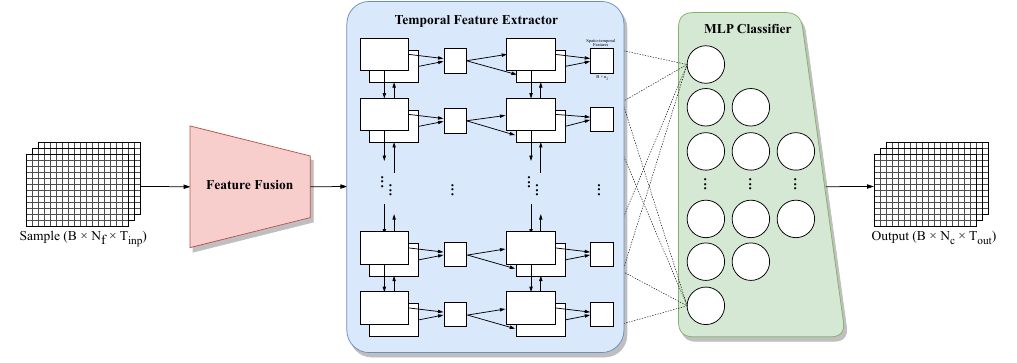}
    \caption{Architecture of our Seq2Seq model}
    \label{fig:model-architecture}
\end{figure*}

\subsubsection{Model Architecture} \label{seq:model-architecture}
As illustrated in Figure \ref{fig:model-architecture}, we employed three main components in our neural network model: a feature fusion (FF) module to combine input features, a temporal feature extractor (TFE) to detect temporal patterns of each sign language gesture, and an MLP classifier to output the predicted gesture using the features obtained from the previous components.

As we had samples of varying lengths in our dataset, we added zero padding to the samples in the same batch. More formally, we formed a $B \times N_f \times T_{inp}$ matrix for each batch, where $B$ is the batch size, $N_f$ is the number of features, and $T_{inp}$ is the length of the longest sequence in the batch. By doing this, we were able to leverage the efficiency of batch gradient descent, thereby accelerating the process. 

The FF component aims to combine the features with each other. It is essential to extract the dependencies between the values of different features before feeding them to the TFE module, which extracts temporal dependencies between them. Moreover, FF extracts local temporal dependencies, which guides TFE in finding more complex patterns. In other words, the raw values of different features in FF are combined with the help of a convolution layer to create higher-level features. For this purpose, we used a 2D convolutional layer with a kernel size of $3 \times 3$ to extract local time-related dependencies while combining different features.

The features generated by the FF component enter the TFE module, which includes two consecutive Bidirectional LSTM (BiLSTM) layers with hidden state sizes $n_{1} = 64$ and $n_{2} = 128$, respectively. Each hidden state unit $h_t$ in BiLSTM contains information from the past and future of the input sequence at time $t$; thus, it helps us distinguish sign language gestures with the same beginning but different endings. Ultimately, each hidden state unit in the last BiLSTM layer yields an output in the form of a matrix with size $B \times n_{2}$, representing the spatio-temporal features distilled from the sequences of each batch.

The final step in our model is when the generated spatio-temporal feature matrices are fed into an MLP classifier to produce the predicted sequence. We used a shared 3-layered perceptron network containing 64, 32, and $N_{classes} + 1$ perceptrons along with the CTC loss function to predict the output label at any time step. Since the CTC loss function requires an extra blank label, we added an extra unit to the last layer, resulting in a total of 17 distinct classes. We further discuss the details of the CTC loss function in Section \ref{sec:CTC-loss}.

To improve the performance of our neural network, we used batch normalization and dropout techniques. Batch normalization contributes significantly to the faster convergence of the network by preventing the covariant shift. Therefore, we used it after the convolution layer of the FF module and between each two layers of the MLP classifier. Later, in Section \ref{sec:ablation}, we will examine the effect of adding batch normalization between different layers of our network. Finally, we placed a dropout layer with a probability of 0.3 after each of the BiLSTM layers. This prevents the model from overfitting and allows it to learn generalizable patterns while training.

\subsubsection{CTC Loss} \label{sec:CTC-loss}
The Connectionist Temporal Classification (CTC) loss function is one of the most popular loss functions in sequence-to-sequence tasks, such as optical character recognition and speech recognition. The main advantage of this function is its capability to align the input sequences to the target sequence, especially in scenarios where this alignment is not one-to-one. In other words, this function is used when the length of input and target sequences do not match, and the exact location of each output event in the input sequence is not specified. Due to the inherent variability in the execution of sign language gestures, using the CTC loss function in SLR can be an effective solution to improve the performance of deep learning models in this field. That is because people execute words at different speeds, and there may be long pauses between movements, which can pose a challenge in accurately recognizing sign language gestures.

CTC introduces a blank symbol and allows the neural network to output repeated occurrences between the actual symbols, thus enabling variable-length output sequences. For example, both (AA-B-C-CC) and (A-B--CC-C) represent the sequence ABCC. During training, the CTC algorithm aligns these output sequences with the target sequences, considering all possible alignments using a dynamic programming approach. Finally, it computes a loss function based on the negative log probability of these alignments, guiding the network to produce output sequences that are likely to match the target sequences.

\subsection{Sequence Alignment}
In this section, we propose a novel method based on ensembling to improve the performance of SLR Seq2Seq models. One way to improve the models' performance in deep learning is to use ensembling. In ensemble methods, several models are trained on the data, and their outputs are used to produce the final prediction. One way to build each of these models is to split the data into $k$ folds and train $k$ different models where each model is trained on $k-1$ folds and validated using the remaining fold (K-fold cross-validation). In the SLR task, however, given a single sample, the outputs of the models trained using K-fold cross-validation could have variable lengths, making it challenging to combine them to generate the final result. To overcome this problem, we suggest using multiple sequence alignment (MSA) algorithms to make the output of the models equal in length and then perform a voting process to produce the final prediction.

\begin{algorithm} [b]
    \caption{Pairwise Global Alignment}
    \label{alg:global-alignment}
    \begin{algorithmic}[1]
    \State \textbf{Input: } Sequences $s_1[0, \ldots, m - 1]$ and $s_2[0, \ldots, n - 1]$
    \State \textbf{Parameters: } $S_{mis}$, $S_{match}$, $S_{gap}$
    \State \textbf{Output: } 2 Aligned Sequences, The Alignment Score
    
    \State Initialize $dp[m+1,n+1]$ with zeros
    \State Initialize $dp[i, 0] = i \times S_{gap}$
    
    \For{$i$ from $1$ to $m$}
        \For{$j$ from $1$ to $n$}
            \If{$s_1[i] = s_2[j]$}
                \State $dp[i, j] = \max\{dp[i-1, j-1] + S_{match}, dp[i-1, j] + S_{gap}, dp[i, j-1] + S_{gap}\}$
            \Else
                \State $dp[i, j] = \max\{dp[i-1, j-1] + S_{mis}, dp[i-1, j] + S_{gap}, dp[i, j-1] + S_{gap}\}$
            \EndIf
        \EndFor
    \EndFor
    
    \State $Aligned_{s1}$, $Aligned_{s2}$ = Backtrack($dp$)
    \State \textbf{return} $Aligned_{s1}$, $Aligned_{s2}$, $dp[m, n]$
    \end{algorithmic}
    \end{algorithm}

Needleman-Wunsch (NW) is an algorithm that uses dynamic programming to compute optimal global alignment (GA) between two strings. As shown in Algorithm \ref{alg:global-alignment}, NW creates a table of size $m+1$ by $n+1$ to align two strings of length $m$ and $n$. Then, the first row and the first column are initialized as follows: 
\begin{equation}
    D_{0, i} = D_{i, 0} = i * S_{gap}
\end{equation}
where $S_{gap}$ is the penalty for creating a gap in one of the sequences. Finally, it fills the table using Eq. \ref{eq:global-dp} and Eq. \ref{eq:global-dp-score}:
\begin{equation}
    D_{ij} = \max
    \begin{cases}
        D_{i-1,j-1} + S(X_i, Y_j) \\
        D_{i-1,j} + S_{gap} \\
        D_{i,j-1} + S_{gap}
    \end{cases}
\label{eq:global-dp}
\end{equation}

\begin{equation}
    S(c_1,c_2) = 
    \begin{cases}
        S_{match} & \text{if } c_1 = c_2 \\
        S_{mis} & \text{o.w}
    \end{cases}    
\label{eq:global-dp-score}
\end{equation}
where $X$ and $Y$ are the input sequences, $S_{match}$ is the score of matching two characters in the sequences, and $S_{mis}$ is the penalty of mismatch between two characters. Then, the final alignment score and the aligned sequences can be obtained by backtracking from the final position $[m,n]$ to the starting position $[0,0]$. Although this method can be extended to align $k$ sequences, its exponential computational complexity prevents us from using it in a real-time system. Therefore, a method with less computational complexity is preferable to align the sequences.

\begin{figure*}[b]
    \includegraphics[width=\textwidth]{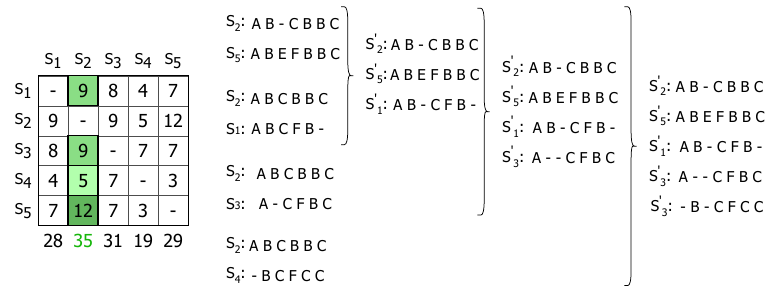}
        \caption{An illustration of how Star Alignment algorithm computes the similarity matrix between 5 sequences ($S_1="ABCFB"$, $S_2="ABCBBC"$, $S_3="ACFBC"$, $S_4="BCFCC"$, $S_5="ABEFBBC"$) and use it to progressively align them. Sequence $S_2$ is designated as the center since the sum of similarities in the second column is the highest. ($S_{match} = 3$, $S_{gap} = -2$, and $S_{mis} = -1$)}
    \label{fig:alignment-star}
\end{figure*}

Star Alignment (SA) is a heuristic method for approximating GA, primarily applied in bioinformatics for aligning DNA sequences. Algorithm \ref{alg:star} shows how SA aligns multiple sequences. First, it calls NW algorithm on each pair of input sequences to compute their alignments and the corresponding similarity scores. Then, the sequence exhibiting the maximum cumulative similarity is designated as the central sequence. Subsequently, other sequences are arranged in descending order based on their similarity to the central sequence. Finally, traversing the sequences in order, some gaps are added to the central string and all previously traversed strings (if necessary) to equalize their lengths, using the alignments calculated in the first step. For instance, Figure \ref{fig:alignment-star} illustrates an step-by-step progress of running SA on five arbitrary sequences.


\begin{algorithm} [t]
    \caption{Star Alignment}
    \label{alg:star}
    \begin{algorithmic}[1]
    \State \textbf{Input:} $k$ sequences $s_0$ to $s_{k-1}$
    \State \textbf{Output:} $k$ Aligned Sequences, The Alignment Score
    
    \State Calculate pairwise alignments and scores between all pairs $(s_i, s_j)$ where $i \neq j$
    \State Choose the sequence $s_c$ as the center, where $c = \arg\max_j \left(\sum_{i=0}^{k-1} \text{Score}(s_j, s_i)\right)$
    \State Sort indices $i \neq c$ in decreasing order with respect to $\text{Score}(s_c, s_i)$
    \For{each index $i$ in the sorted indices}
        \State Utilize the alignments between $s_c$ and $s_i$ to update the gap locations of the aligned sequences $s'_c$ and $s'_j$ where $j \leq i$
    \EndFor
    \State \textbf{return} The Aligned Sequences $s'_i$, Total Alignment Score
    \end{algorithmic}
    \end{algorithm}

\begin{figure*}[b]
\includegraphics[width=\textwidth]{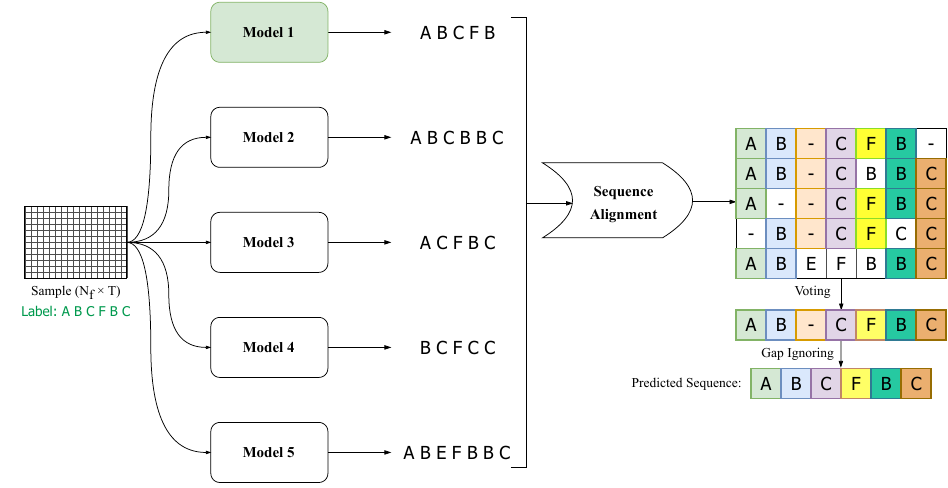}
    \caption{An example of the execution of our ensembling algorithm. The outputs produced by all models are aligned using Star Alignment, after which a voting process is performed to obtain the most probable gesture (denoted by characters A to F) at each position. Finally, the gaps are ignored, and the final result is generated. This example is a situation when the best model (Model 1) can not completely predict the ground truth sequence, but the ensembling method helps the system to generate it successfully.}
    \label{fig:alignment-voting}
\end{figure*}

We utilize SA along with K-fold cross-validation to enhance the accuracy of our system. Algorithm \ref{alg:voting} provides a detailed overview of our method. In the training phase, we use 5-fold cross-validation to train five distinct models on Dataset1-3, and when testing, we use those models to predict five sequences $s_0$ to $s_4$ for each input sample. Next, we use the predicted sequences as the inputs of the SA algorithm to obtain aligned sequences $s'_0$ to $s'_4$. The final step includes voting at each time step $t$ of the final alignments. In other words, the most prevalent character at $t$ is added to the final result unless the character is a gap. It should be noted that before using SA, it is necessary to map the sign language words to arbitrary characters. Therefore, we restore the words using inverse mapping after the final voting. Figure \ref{fig:alignment-voting} depicts the execution of the explained process.


\begin{algorithm} [t]
    \caption{Our Ensembling Algorithm}
    \label{alg:voting}
    \begin{algorithmic}[1]
    \State \textbf{Input:} Sample $x$, Mapping T from the chosen sign language words to arbitrary characters
    \State \textbf{Output:} The Predicted Sequence
    
    \State Train 5 models via 5-fold cross-validation 
    \State Predict the outputs ($s_0$ to $s_4$) for the input $x$ 
    \State Replace each character $ch$ in $s_i$ with $T[ch]$
    \State Pass sequences $s_i$ to Star Aignment to get sequences $s'_i$
    \For{$t$ from 0 to $s'_0.length - 1$}
        \State $out = out + \text{Vote}(s_0[t], ..., s_4[t])$
    \EndFor
    \State Remove gaps from $out$
    \State Use the inverse mapping of $T$ to update $output$
    \State \textbf{return} $out$
    \end{algorithmic}
    \end{algorithm}

The intuition behind our model relies on the fact that the models trained on distinct folds tend to learn different patterns, especially when the dataset is not very large. That is, their accuracy in detecting different classes may vary, and each could manage to predict segments of the ground truth sequence. As a result, by aligning their predictions and voting between them, the final result is expected to have a lower error rate than the prediction of the best one. As will be explained in Section \ref{sec:ensembling-performance}, this method does not always improve the accuracy. This situation occurs when the best model has much higher accuracy than others or when all models do not have enough accuracy, so the performance drops by voting between the predictions. However, the results depict that, on average, the method could significantly improve the performance of the system. Furthermore, since our ensembling technique does not depend on any linguistic feature, it can be adapted for studies on other variations of sign language or language-independent sequence-to-sequence tasks such as gesture recognition.


\section{Experiment Setup and Results} \label{sec:eval}
In this section, we examine the performance of the proposed model and our ensembling algorithm. We use a subject-independent approach in all experiments unless where it is mentioned explicitly. By using leave-one-subject-out cross-validation, we train each model on the data of 4 subjects and test it with the data of the remaining subject. In the training phase, we use Dataset1-3, mentioned in Section \ref{sec:dataset}, which includes sentences with up to three words. In addition, we use the 5-fold cross-validation approach to apply our ensembling algorithm, where each fold contains 20\% of the data. Also, we use the AdamW optimizer with a batch size of 9 and learning rate scheduling in 3 steps to optimize the cost function as much as possible. On the other hand, when testing, our experiment settings involve both Dataset1-3 and Dataset4-8, enabling us to evaluate user-independent generalization as well as generalization in longer sentences, respectively. Furthermore, in the testing phase, we assigns scores to gap penalties, mismatches, and matches as follows: $S_{gap} = -1$, $S_{mis} = -1$, and $S_{match} = 0$. This configuration was determined to be optimal for our recognition task through rigorous experimentation, resulting in the best performance.

\subsection{Evaluation Metrics}
We use the following four metrics to analyze the performance of our models: \\
\textbf{Sequence Length Accuracy (SLAcc):} It depicts the percentage of sequences whose lengths are accurately predicted by the model relative to the total number of sequences. \\
\textbf{Sequence Accuracy (SAcc):} It indicates the percentage of sequences that exactly match their ground truth out of the total number of sequences.
Since achieving high results in this metric is quite challenging, mainly when the dataset contains lengthy sequences, many previously conducted studies have not reported their results based on this metric. Despite this, we report our results to show the effectiveness of our ensembling approach in correcting the mispredicted sequences. \\
\textbf{Word Accuracy (WAcc):} This metric is defined based on another metric called Word Error Rate (WER), which is used in tasks where the length of the predicted sentences could be different from their actual length. To calculate WER for a given pair of ground truth and predicted sequences, we use the following equation:
\begin{equation}
    WER = \frac{S + D + I}{L}\times100
\end{equation}
where $L$ is the length of the ground truth sequence, $S$ is the number of substitutions, $I$ is the number of insertions, and $D$ is the number of deletions needed to convert the ground truth sequence to the predicted sequence. In fact, the numerator is the widely used Levenshtein distance, which calculates the minimum number of single-character edits needed to transform one sequence into another. In our task, each individual sign language gesture in predicted sequences corresponds to a character in the Levenshtein distance calculation. Then, we compute word accuracy to show the similarity of the predicted sequence to the corresponding ground truth:
\begin{equation}
    \text{WAcc} = 100 - \text{WER}
\end{equation}
Finally, we calculate total word accuracy as follows:
\begin{equation}
    \text{Total}\ \text{WAcc} = \frac{\sum \text{WAcc}_i}{N}
\end{equation}
where $WAcc_i$ is the word accuracy for the $i$-th sample in the dataset, and $N$ is the total number of samples. \\
\textbf{Weighted Word Accuracy (WWAcc):} Instead of computing a simple average over the obtained word accuracies, we can take a weighted average to amplify the impact of lengthy sentences in the final result. Thus, we can calculate weighted word accuracy using the following equation:
\begin{equation}
    \text{WWAcc} = \frac{\sum \text{WAcc}_i . L_i}{\sum L_i}
\end{equation}
Contrasting these results with the previous metric can showcase the extendability of our approach to longer sentences.

\subsection{Ablation Study} \label{sec:ablation}
As mentioned in Section \ref{seq:model-architecture}, our model includes a $3 \times 3$ convolutional layer, two BiLSTM layers, and an MLP classifier with batch normalization between any two layers. To show the impact of different components in the proposed architecture, we remove or change different parts of the model. Table \ref{tab:ablation} shows the results of this work without applying our ensembling method. In the last column of this table, the accuracies obtained from the performance of the proposed model are shown. Other columns depict the amount of change in the model's performance after removing or changing one of the model's components. It should be noted that the numbers shown are the average test results of the model across different subjects while having trained on a specific fold of Dataset1-3 without applying our ensembling method.

\begin{table}[b]
    \resizebox{0.7\textwidth}{!}{%
    \begin{tabular}{ccccccc|>{\columncolor{green!20}}c}
        \toprule
        & \multicolumn{2}{c}{\textbf{CNN}} & \multicolumn{1}{c}{\textbf{LSTM}} & \multicolumn{1}{c}{\textbf{Bidirectional}} & \multicolumn{2}{c|}{\textbf{Batch Norm}} & \multicolumn{1}{c}{\textbf{Our Model}} \\
        \cmidrule(){2-3} \cmidrule(){6-7}    
        & \multicolumn{1}{c}{No} & \multicolumn{1}{c}{$5 \times 5$} & \multicolumn{1}{c}{1} & \multicolumn{1}{c}{No} & \multicolumn{1}{c}{0} & \multicolumn{1}{c|}{1} \\
        \midrule
        \textbf{SLAcc} & -4.92 & -3.05 & -4.22 & -12.33 & -1.50 & -1.86 & \cellcolor{green!20}91.47 \\
        \textbf{SAcc} & -15.12 & -2.51 & -1.55 & -12.09 & -4.41 & -0.23 & \cellcolor{green!20}81.69 \\
        \textbf{WAcc} & -8.91 & -2.07 & -1.24 & -6.97 & -2.71 & -0.39 & \cellcolor{green!20}90.39 \\
        \textbf{WWAcc} & -8.71 & -1.38 & -0.70 & -6.01 & -2.56 & -0.11 & \cellcolor{green!20}90.85 \\
        \textbf{Average} & -9.41 & -2.25 & -1.93 & -9.35 & -2.79 & -0.65 & \cellcolor{green!20}88.60 \\
        \bottomrule
    \end{tabular}}%
    \caption{The percentage of performance degradation of models resulting from the deletion or modification of each component within our proposed architecture.}
    \label{tab:ablation}
\end{table}

The model achieves a WAcc of 90.39\% and a WWAcc of 90.85\%, resulting in an exact match ratio of 81.69\%. Moreover, the model is able to produce sequences with the same length as the ground truth in 91.47\% of samples. Due to the negativeness of other numbers in the table, applying any of the mentioned changes in the architecture causes the performance of the model to drop. Avoiding the use of the convolutional or BiLSTM layers has a significant impact on the model's performance, which respectively indicates the high importance of the Feature Fusion component and the role of the bidirectional LSTM in detecting gestures with the same opening movements. Also, using just one LSTM layer instead of two layers or using a $5 \times 5$ kernel for the convolution layer reduces the performance. The former is due to the fact that using two layers of LSTM can help us detect more complex patterns in the data, and the latter is because the $3 \times 3$ kernel allows features to be combined in more ways. Finally, the use of batch normalization between each pair of perceptron layers makes the length of the predicted sentences much closer to their actual value, which ultimately increases the accuracy of the model.

\begin{table}[b]
     \resizebox{0.9\textwidth}{!}{%
     \centering
      \begin{tabular}{{P{1cm}P{1cm}P{2cm}P{2cm}P{2cm}P{2cm}P{2cm}P{2cm}}}
     \toprule
          &   &                                          \textbf{Subject 1} &                                          \textbf{Subject 2} &                                          \textbf{Subject 3} &                                         \textbf{Subject 4} &                                          \textbf{Subject 5} &                                            \textbf{Average} \\
          \textbf{Metric} & \textbf{Model} &                                                    &                                                    &                                                    &                                                   &                                                    &                                                    \\
     \midrule
     \multirow{2}{*}{\textbf{SLAcc}} & 1 &   \textcolor{ForestGreen}{93.22} /  \textcolor{black}{90.58} &     \textcolor{red}{91.54} /  \textcolor{black}{95.34} &   \textcolor{ForestGreen}{96.01} /  \textcolor{black}{95.51} &     \textcolor{red}{95.45} /  \textcolor{blue}{97.56} &      \textcolor{blue}{97.55} /  \textcolor{red}{96.73} &    \textcolor{black}{94.79} /  \textcolor{blue}{95.15} \\
          & 2 &      \textcolor{red}{86.94} /  \textcolor{blue}{90.91} &    \textcolor{blue}{96.03} /  \textcolor{ForestGreen}{97.58} &      \textcolor{blue}{95.85} /  \textcolor{red}{93.69} &   \textcolor{black}{96.10} /  \textcolor{ForestGreen}{97.89} &   \textcolor{black}{96.73} /  \textcolor{ForestGreen}{98.85} &     \textcolor{red}{94.33} /  \textcolor{ForestGreen}{95.78} \\
     \cline{1-8}
     \multirow{2}{*}{\textbf{SAcc}} & 1 &   \textcolor{black}{82.31} /  \textcolor{ForestGreen}{82.98} &      \textcolor{red}{86.87} /  \textcolor{blue}{89.64} &     \textcolor{blue}{89.04} /  \textcolor{ForestGreen}{90.70} &    \textcolor{red}{83.77} /  \textcolor{black}{85.39} &     \textcolor{black}{91.82} /  \textcolor{red}{91.16} &   \textcolor{black}{86.76} /  \textcolor{ForestGreen}{87.95} \\
          & 2 &      \textcolor{red}{78.02} /  \textcolor{blue}{82.81} &   \textcolor{black}{89.12} /  \textcolor{ForestGreen}{91.88} &     \textcolor{black}{86.38} /  \textcolor{red}{85.71} &    \textcolor{blue}{87.01} /  \textcolor{ForestGreen}{87.50} &    \textcolor{blue}{91.82} /  \textcolor{ForestGreen}{91.82} &      \textcolor{red}{86.46} /  \textcolor{blue}{87.92} \\
     \cline{1-8}
     \multirow{2}{*}{\textbf{WAcc}} & 1 &    \textcolor{black}{91.32} /  \textcolor{blue}{91.76} &      \textcolor{red}{93.09} /  \textcolor{blue}{94.39} &     \textcolor{blue}{94.38} /  \textcolor{ForestGreen}{95.60} &    \textcolor{red}{90.34} /  \textcolor{black}{91.31} &      \textcolor{red}{95.34} /  \textcolor{blue}{95.74} &   \textcolor{black}{92.89} /  \textcolor{ForestGreen}{93.75} \\
          & 2 &     \textcolor{red}{89.39} /  \textcolor{ForestGreen}{91.76} &   \textcolor{black}{94.27} /  \textcolor{ForestGreen}{95.54} &     \textcolor{black}{93.27} /  \textcolor{red}{92.91} &   \textcolor{blue}{91.75} /  \textcolor{ForestGreen}{92.32} &   \textcolor{ForestGreen}{95.74} /  \textcolor{black}{95.72} &      \textcolor{red}{92.87} /  \textcolor{blue}{93.63} \\
     \cline{1-8}
     \multirow{2}{*}{\textbf{WWAcc}} & 1 &    \textcolor{black}{91.10} /  \textcolor{ForestGreen}{91.48} &       \textcolor{red}{93.68} /  \textcolor{blue}{95.00} &    \textcolor{blue}{93.82} /  \textcolor{ForestGreen}{94.94} &    \textcolor{red}{92.22} /  \textcolor{black}{93.13} &     \textcolor{black}{95.94} /  \textcolor{red}{95.72} &   \textcolor{black}{93.35} /  \textcolor{ForestGreen}{94.04} \\
          & 2 &      \textcolor{red}{89.43} /  \textcolor{blue}{91.33} &   \textcolor{black}{94.77} /  \textcolor{ForestGreen}{96.02} &     \textcolor{black}{92.45} /  \textcolor{red}{92.37} &   \textcolor{blue}{93.66} /  \textcolor{ForestGreen}{94.18} &    \textcolor{ForestGreen}{96.17} /  \textcolor{blue}{96.02} &      \textcolor{red}{93.29} /  \textcolor{blue}{93.97} \\
     \bottomrule
     \end{tabular}}
\caption{Evaluation of models based on the subject-independent approach on Dataset1-3. Each column (except the average column) illustrates the results obtained by testing the model for the corresponding subject while having trained on the data from other subjects. The numbers in the first and second rows of each cell represent models $M_1$ and $M_2$, respectively. The numbers on the left indicate the performance before ensembling, while the numbers on the right indicate the performance after ensembling.}
\label{tab:seq-alignment-dataset1-3}
\end{table}

\subsection{Ensembling Performance} \label{sec:ensembling-performance}
In this section, we examine the performance of the proposed ensembling method and its effectiveness in improving accuracy. According to Table \ref{tab:ablation}, since the results of the model that uses one batch normalization layer ($M_1$) are, on average, close to our proposed model ($M_2$), we will compare these models in our experiments. Also, in addition to Dataset1-3, we use Dataset4-8 to evaluate the performance of the model in longer sentences.

The tables in this section show the performance of $M_1$ and $M_2$ before and after applying ensembling on different datasets and with different approaches. The models related to Table \ref{tab:seq-alignment-dataset1-3} and Table \ref{tab:seq-alignment-dataset4-8-splitted} were trained on the data of 4 subjects from Dataset1-3 and tested on the data of the remaining subject from Dataset1-3 and Dataset4-8, respectively. On the other hand, Table \ref{tab:seq-alignment-dataset4-8-combined} shows the performance of the models by being trained on the data of 4 subjects from Dataset1-3 and tested on all the data of Dataset4-8; therefore, the results in this table are not based on the subject-independent approach. It should be noted that in each cell, the numbers on the left indicate the performance of the best model among the five models trained using 5-fold cross-validation before applying ensembling. In contrast, the numbers on the right show the performance of the ensembled models.

Although some cells in the tables show minor degradation, in the average column, either one of the ensembled models outperforms others across all 12 cases. Notably, in eight cases, the top two ranks are also secured by ensembled models. In the remaining four cases, $M_2$ alone outperforms the ensembled model of $M_1$ due to its inherent excellence. By assigning numerical values from 1 to 4 to the spectrum of colors, ranging from red to green, and substituting them with the corresponding values in each cell, we can compute an average rank for each model. This provides an overall view of the effect of the ensembling method. The result of this analysis is represented in Table \ref{tab:average-ranks-in-tables}. It illustrates a similar pattern in the results of Table \ref{tab:seq-alignment-dataset1-3} and Table \ref{tab:seq-alignment-dataset4-8-combined}, where the ensembled models of $M_2$ and $M_1$ are in the first and second place, respectively. On the contrary, the results of Table \ref{tab:seq-alignment-dataset4-8-splitted} show $M_2$ and its ensemble yielding the best results. This observation indicates that the use of batch normalization between all layers in the classifier improves the performance of $M_2$, such that it can perform even better than the ensembled version of $M_1$. In other words, batch normalization enables the model to generalize better in longer unseen sequences when the subject-independent approach is used.
\begin{table}[t]
     \resizebox{0.9\textwidth}{!}{%
     \begin{tabular}{{P{1cm}P{1cm}P{2cm}P{2cm}P{2cm}P{2cm}P{2cm}P{2cm}}}
     \toprule
          &   &                                         \textbf{Subject 1} &                                         \textbf{Subject 2} &                                         \textbf{Subject 3} &                                         \textbf{Subject 4} &                                         \textbf{Subject 5} &                                           \textbf{Average} \\
     \textbf{Metric} & \textbf{Model} &                                                   &                                                   &                                                   &                                                   &                                                   &                                                   \\
     \midrule
     \multirow{2}{*}{\textbf{SLAcc}} & 1 &     \textcolor{red}{55.00} / \textcolor{black}{75.00} &      \textcolor{red}{75.00} / \textcolor{blue}{90.00} &   \textcolor{red}{100.0} / \textcolor{black}{100.0} &    \textcolor{red}{95.00} / \textcolor{black}{100.0} &     \textcolor{red}{85.00} / \textcolor{black}{85.00} &     \textcolor{red}{82.00} / \textcolor{black}{90.00} \\
          & 2 &    \textcolor{blue}{90.00} / \textcolor{ForestGreen}{90.00} &   \textcolor{black}{85.00} / \textcolor{ForestGreen}{90.00} &  \textcolor{blue}{100.0} / \textcolor{ForestGreen}{100.0} &  \textcolor{blue}{100.0} / \textcolor{ForestGreen}{100.0} &    \textcolor{blue}{95.00} / \textcolor{ForestGreen}{95.00} &    \textcolor{blue}{94.00} / \textcolor{ForestGreen}{95.00} \\
     \cline{1-8}
     \multirow{2}{*}{\textbf{SAcc}} & 1 &     \textcolor{red}{45.00} / \textcolor{black}{60.00} &   \textcolor{black}{55.00} / \textcolor{ForestGreen}{60.00} &     \textcolor{red}{90.00} / \textcolor{black}{95.00} &     \textcolor{red}{85.00} / \textcolor{black}{90.00} &      \textcolor{red}{55.00} / \textcolor{blue}{70.00} &     \textcolor{red}{66.00} / \textcolor{black}{75.00} \\
          & 2 &    \textcolor{blue}{75.00} / \textcolor{ForestGreen}{75.00} &      \textcolor{red}{45.00} / \textcolor{blue}{55.00} &    \textcolor{blue}{95.00} / \textcolor{ForestGreen}{95.00} &    \textcolor{blue}{95.00} / \textcolor{ForestGreen}{95.00} &   \textcolor{black}{65.00} / \textcolor{ForestGreen}{75.00} &    \textcolor{blue}{75.00} / \textcolor{ForestGreen}{79.00} \\
     \cline{1-8}
     \multirow{2}{*}{\textbf{WAcc}} & 1 &    \textcolor{red}{82.04} / \textcolor{black}{88.90} &   \textcolor{blue}{90.20} / \textcolor{ForestGreen}{90.73} &    \textcolor{red}{94.38} / \textcolor{black}{95.00} &   \textcolor{red}{96.92} / \textcolor{black}{98.33} &   \textcolor{red}{89.49} / \textcolor{black}{93.87} &   \textcolor{red}{90.61} / \textcolor{black}{93.37} \\
          & 2 &  \textcolor{ForestGreen}{94.88} / \textcolor{blue}{93.88} &    \textcolor{red}{86.68} / \textcolor{black}{88.90} &    \textcolor{blue}{95.00} / \textcolor{ForestGreen}{95.00} &  \textcolor{blue}{99.17} / \textcolor{ForestGreen}{99.17} &  \textcolor{blue}{94.61} / \textcolor{ForestGreen}{95.95} &  \textcolor{blue}{94.07} / \textcolor{ForestGreen}{94.58} \\
     \cline{1-8}
     \multirow{2}{*}{\textbf{WWAcc}} & 1 &   \textcolor{red}{83.33} / \textcolor{black}{90.83} &  \textcolor{blue}{90.83} / \textcolor{ForestGreen}{90.83} &   \textcolor{red}{93.33} / \textcolor{black}{94.17} &    \textcolor{red}{97.50} / \textcolor{black}{98.33} &   \textcolor{red}{90.83} / \textcolor{black}{94.17} &   \textcolor{red}{91.16} / \textcolor{black}{93.67} \\
          & 2 &   \textcolor{ForestGreen}{95.83} / \textcolor{blue}{95.00} &    \textcolor{red}{87.50} / \textcolor{black}{89.17} &  \textcolor{blue}{94.17} / \textcolor{ForestGreen}{94.17} &  \textcolor{blue}{99.17} / \textcolor{ForestGreen}{99.17} &  \textcolor{blue}{94.17} / \textcolor{ForestGreen}{95.83} &  \textcolor{blue}{94.17} / \textcolor{ForestGreen}{94.67} \\
     \bottomrule
     \end{tabular}}
\caption{Evaluation of models based on the subject-independent approach on Dataset4-8}
\label{tab:seq-alignment-dataset4-8-splitted}
\end{table}

\begin{table}[t]
     \resizebox{0.9\textwidth}{!}{%
     \begin{tabular}{{P{1cm}P{1cm}P{2cm}P{2cm}P{2cm}P{2cm}P{2cm}P{2cm}}}
     \toprule
          &   &                                        \textbf{Subject 1} &                                         \textbf{Subject 2} &                                          \textbf{Subject 3} &                                          \textbf{Subject 4} &                                          \textbf{Subject 5} &                                            \textbf{Average} \\
          \textbf{Metric} & \textbf{Model} &                                                  &                                                   &                                                    &                                                    &                                                    &                                                    \\
     \midrule
     \multirow{2}{*}{\textbf{SLAcc}} & 1 &     \textcolor{red}{86.00} / \textcolor{blue}{91.00} &     \textcolor{red}{84.00} / \textcolor{ForestGreen}{95.00} &     \textcolor{black}{89.00} / \textcolor{blue}{92.00} &       \textcolor{red}{84.00} / \textcolor{blue}{94.00} &       \textcolor{red}{87.00} / \textcolor{blue}{94.00} &       \textcolor{red}{86.00} / \textcolor{blue}{93.20} \\
          & 2 &  \textcolor{black}{90.00} / \textcolor{ForestGreen}{94.00} &    \textcolor{black}{88.00} / \textcolor{blue}{94.00} &      \textcolor{red}{87.00} / \textcolor{ForestGreen}{95.00} &    \textcolor{black}{93.00} / \textcolor{ForestGreen}{96.00} &    \textcolor{black}{91.00} / \textcolor{ForestGreen}{97.00} &    \textcolor{black}{89.80} / \textcolor{ForestGreen}{95.20} \\
     \cline{1-8}
     \multirow{2}{*}{\textbf{SAcc}} & 1 &     \textcolor{red}{74.00} / \textcolor{blue}{81.00} &     \textcolor{red}{73.00} / \textcolor{ForestGreen}{86.00} &      \textcolor{red}{79.00} / \textcolor{ForestGreen}{90.00} &       \textcolor{red}{75.00} / \textcolor{blue}{88.00} &       \textcolor{red}{73.00} / \textcolor{blue}{84.00} &       \textcolor{red}{74.80} / \textcolor{blue}{85.80} \\
          & 2 &  \textcolor{black}{79.00} / \textcolor{ForestGreen}{86.00} &    \textcolor{black}{77.00} / \textcolor{blue}{84.00} &     \textcolor{black}{79.00} / \textcolor{blue}{88.00} &    \textcolor{black}{87.00} / \textcolor{ForestGreen}{92.00} &    \textcolor{black}{82.00} / \textcolor{ForestGreen}{90.00} &    \textcolor{black}{80.80} / \textcolor{ForestGreen}{88.00} \\
     \cline{1-8}
     \multirow{2}{*}{\textbf{WAcc}} & 1 &  \textcolor{red}{92.46} / \textcolor{black}{95.21} &    \textcolor{red}{92.40} / \textcolor{ForestGreen}{96.15} &   \textcolor{black}{95.30} / \textcolor{ForestGreen}{97.05} &     \textcolor{red}{94.27} / \textcolor{blue}{97.03} &     \textcolor{red}{93.21} / \textcolor{blue}{96.08} &      \textcolor{red}{93.53} / \textcolor{blue}{96.30} \\
          & 2 &  \textcolor{blue}{95.55} / \textcolor{ForestGreen}{96.50} &  \textcolor{black}{94.14} / \textcolor{blue}{95.28} &     \textcolor{red}{94.78} / \textcolor{blue}{96.66} &   \textcolor{black}{96.46} / \textcolor{ForestGreen}{97.70} &  \textcolor{black}{95.99} / \textcolor{ForestGreen}{97.37} &   \textcolor{black}{95.38} / \textcolor{ForestGreen}{96.70} \\
     \cline{1-8}
     \multirow{2}{*}{\textbf{WWAcc}} & 1 &  \textcolor{red}{92.17} / \textcolor{black}{95.33} &   \textcolor{red}{92.83} / \textcolor{ForestGreen}{96.17} &  \textcolor{black}{95.33} / \textcolor{ForestGreen}{97.17} &      \textcolor{red}{94.33} / \textcolor{blue}{97.00} &     \textcolor{red}{93.67} / \textcolor{blue}{96.17} &     \textcolor{red}{93.67} / \textcolor{blue}{96.37} \\
          & 2 &   \textcolor{blue}{95.50} / \textcolor{ForestGreen}{96.50} &   \textcolor{black}{94.17} / \textcolor{blue}{95.50} &      \textcolor{red}{95.00} / \textcolor{blue}{96.83} &  \textcolor{black}{96.67} / \textcolor{ForestGreen}{97.67} &   \textcolor{black}{96.00} / \textcolor{ForestGreen}{97.33} &  \textcolor{black}{95.47} / \textcolor{ForestGreen}{96.77} \\
     \bottomrule
     \end{tabular}}
\caption{Evaluation of models based on the subject-dependent approach on Dataset4-8. Each column (except the average column) illustrates the results obtained by testing the model for the corresponding subject while having trained on the data from all subjects.}
\label{tab:seq-alignment-dataset4-8-combined}
\end{table}

So far, we have demonstrated that despite causing minor degradations in certain metrics for specific subjects, the application of ensembling, on average, improves the overall performance. Now, we will examine the extent of this impact. Table \ref{tab:seq-alignment-dataset1-3} indicates that ensembling boosts both WAcc and SLAcc, resulting in 1.19\% and 1.46\% improvement in SAcc of $M1$ and $M2$, respectively. Table \ref{tab:seq-alignment-dataset4-8-splitted} reveals that ensembling effectively addresses $M_1$'s weakness in recognizing sequences with 4 to 8 words, bringing it closer to $M_2$'s performance with a 2.76\% and 0.51\% increase in WAcc for $M_1$ and $M_2$, respectively. This translates to a remarkable 9.00\% and 4.00\% enhancement in recognizing complete sequences for each model. Moreover, the models achieved slightly higher results in terms of WWAcc, indicating that they are capable of maintaining their accuracy while dealing with longer sentences. The results of Table \ref{tab:seq-alignment-dataset4-8-combined} show that the impact of ensembling is even higher while using the subject-dependent approach. In that experiment, $M_1$ and $M_2$ witnessed a 2.77\% and 1.32\% increase in WAcc, resulting in an 11.00\% and 7.20\% improvement in recognizing complete sequences, respectively. Moreover, our method significantly improved the ability of models to predict the length of output sequence, resulting in 7.20\% and 5.40\% improvement, respectively.

To wrap up, our experiments in this section show two main points. First, a better generalization in longer sequences is the underlying cause for using batch normalization between all layers in the classifier. Second, using our ensembling approach yields considerable improvements in both models, making it a suitable choice for our SLR task.

\begin{table}[t]
    \resizebox{0.6\textwidth}{!}{%
     \begin{tabular}{{P{1.75cm}P{1.75cm}P{1.75cm}P{1.75cm}P{1.75cm}}}
     \toprule
          & \textbf{Model} &                                          \textbf{Table 2} &                                          \textbf{Table 3} &                                          \textbf{Table 4} \\
     \midrule
     \multirow{2}{*}{\textbf{Average ranks}}  & 1 &
      \textcolor{red}{1.96} /  \textcolor{blue}{2.83} &       \textcolor{red}{1.21} /  \textcolor{black}{2.33} &      \textcolor{red}{1.13} /  \textcolor{blue}{3.21} \\ & 2 &    \textcolor{black}{2.08} /  \textcolor{ForestGreen}{3.13} &        \textcolor{blue}{2.75} /  \textcolor{ForestGreen}{3.71} &        \textcolor{black}{1.95} /  \textcolor{ForestGreen}{3.71} \\
     \bottomrule
     \end{tabular}}
\caption{The average ranking of each model based on the results of experiments. The ranks range from 1 to 4 corresponding to models' performance in ascending order.}
\label{tab:average-ranks-in-tables}
\end{table}

\section{Limitations and Future Work} \label{sec:limit}

\subsection{Neglecting Non-manual Markers}
Although PenSLR is capable of capturing diverse hand movements and finger positions, it lacks the ability to detect non-manual markers, such as facial expressions and head movements. As previously mentioned, non-manual markers play a crucial role in all sign language variations. In fact, some gestures may share similar hand movements and finger configurations, yet differ in subtle facial expressions. Therefore, incorporating non-manual markers not only expands the range of glosses but also enhances the system's ability to differentiate between gestures with similar manual markers. In the future, we could manipulate and improve our glove-based design so that we can take into account these non-manual markers. To achieve this goal, one promising approach involves employing stretchable sensors attached to the face, as suggested in \cite{10-zhou2020sign}, or harnessing earbud signals, as explored in SmartASL \cite{16-jin2023smartasl}.

\subsection{Limited Dataset}
As explained in Section \ref{sec:dataset}, we collected a dataset of over 3000 samples from 16 PSL glosses. Although our dataset stands as the largest within the PSL domain, the PSL lexicon extends beyond this limited scope, and for effective communication among sign language users, a broader vocabulary is essential. Therefore, it is crucial to prioritize the expansion of our dataset to
include a more extensive array of PSL signs. This will facilitate more comprehensive communication within the PSL community. Additionally, collaborating with the Persian Deaf Community Association, particularly in the data gathering process, could help us to improve the quality of our dataset and tailor it to real-world applications.

\subsection{Ensembling Limitations}
Our ensembling algorithm does not take into account any linguistic relationship between the predicted signs. Although this counts as a benefit in language-independent tasks, such as gesture recognition, the situation is different when linguistic features are decisive. For instance, in sign language translation, where the goal is to convert sign language glosses into natural language text, our algorithm may fall short in preserving lexical and grammatical correctness. This limitation arises because the use of sequence alignment can disrupt the coherence of sentence structures. However, to overcome this issue, a language model can be used to revise the output of ensembling to preserve the sentence structure.

\section{Conclusion} \label{sec:conc}

In this paper, we propose PenSLR, the first Persian sign language recognition system capable of detecting sign language sentences in an end-to-end manner. To achieve this, we design a cost-effective glove-based system that can capture a wide range of manual markers. Moreover, we construct and publish the largest PSL dataset to date, comprising more than 3000 samples from 16 commonly used signs. To tackle the recognition task, we deploy a CRNN architecture coupled with the CTC loss function, facilitating the generation of variable-length outputs based on input signals. Furthermore, we introduce a novel ensembling technique based on sequence alignment and utilize multiple instances of our model to generate an improved output sequence. PenSLR achieves an impressive 94.07\% word-level accuracy while testing on longer unseen sentences using the subject-independent approach. Notably, when employing our ensembling scheme, this represents a 0.51\% enhancement over the performance achieved solely with the CRNN model, leading to a 94.58\% word-level accuracy. This effect of the ensembling is even higher when a subject-dependent approach is used, where the test data belongs to the same subjects as the training data.

\begin{acks}
We would like to extend a special thanks to the five volunteers who generously donated their time and effort to help us record and collect the necessary data. Their unwavering commitment was crucial to the success of our project. 


\end{acks}

\bibliographystyle{ACM-Reference-Format}
\bibliography{sample-manuscript}

\appendix









\end{document}